\documentclass[aps,pre,reprint]{revtex4-1}
\usepackage[T1]{fontenc}
\usepackage{amsmath}
\usepackage[latin1]{inputenc}
\usepackage{graphicx}
\usepackage[normalem]{ulem}

\usepackage{epsfig}

\usepackage{color}

\begin{document}
\title{Temperature (de)activated patchy colloidal particles}

\author{Daniel de las Heras}
\email{delasheras.daniel@gmail.com}
\affiliation{Theoretische Physik II, Physikalisches Institut, 
  Universit{\"a}t Bayreuth, D-95440 Bayreuth, Germany}

\author{Margarida M. Telo da Gama}
\email{mmgama@fc.ul.pt}
\affiliation{Departamento de F\'{\i}sica e Centro de F\'{\i}sica Te\'orica e Computacional, Faculdade de Ci\^encias, Universidade de Lisboa, Campo Grande, P-1749-016, Lisbon, Portugal}

\date{\today}
\begin{abstract}
We present a new model of patchy particles in which the interaction sites can be activated or deactivated by varying the temperature of the system. We study the thermodynamics of the system by means of Wertheim's first order perturbation theory, and use Flory-Stockmayer theory of polymerization to analyse the percolation threshold. We find a very rich phase behaviour including lower critical points and reentrant percolation. 

\end{abstract}

\maketitle

\section {Introduction}
Patchy colloids interact via a very directional, valence limited potential. To achieve this, the surface of the colloids is patterned with bonding sites or patches. By playing with the number, type, and specific shape of the patches it is possible to find a remarkable variety of new phenomenology not present in colloidal particles interacting via isotropic potentials. Examples are empty liquids \cite{PhysRevLett.97.168301,C0SM01493A,emptyexp}, reentrant networks \cite{PhysRevLett.106.085703,Tav2}, and self-assembly into complex structures \cite{Granick1,Sciortino2,0953-8984-25-19-193101}. 

Patchy colloids form reversible bonds between bonding sites if two sites are close enough. Typically, the patches are always activated in the sense that they are always available to form bonds with the surrounding patches. It is possible, however, to design patchy colloids in such a way that the patches can be deactivated under some circumstances. For example, in \cite{roldan2013gelling} Sciortino et al. studied theoretically a mixture of large patchy colloids with four patches of type A and small colloids with only one patch of type B. Patches of type A can form bonds with patches of both types A and B. Bonds between two patches of type B are not allowed. The AB bond is stronger than the AA bond. As a result at low temperatures the system maximizes bonds of type AB, which in practice is equivalent to breaking the connectivity of the system via deactivation of bonds AA. At intermediate temperatures the entropy plays a major role and the system forms a gel-like structure. In a recent work, Roth \cite{r2014ion} et al. used patchy particles to model proteins. The patch-patch interaction is mediated by ions that activate the patches. A bond between patches is possible only if one, and only one, of the patches is occupied by an ion. The phase diagram shows a reentrant condensation as function of the concentration of ions since all patches are deactivated if the concentration of ions is either very low or very high.

A possible route to experimentally fabricate bonding sites activated by temperature consists on using colloids with patches made of complementary strands of DNA \cite{Eiser1}. The patches are active only at temperatures below the melting temperature of the DNA. The melting temperature depends, among other factors, on the DNA sequence. Therefore, by using patches coated with distinct DNA sequences one might fabricate complex patchy colloids with distinct activation temperatures. This method has been used in \cite{Eiser2} to design a multistep self-assembly in isotropically DNA-coated colloids. Also, in~\cite{Re2} a binary system of DNA-coated colloids was investigated using Monte Carlo simulation. Each species is coated with two different DNA strands that are designed such that the binary system exhibits reentrant melting. That is, there is a fluid state at high and low temperatures, and a crystal at intermediate temperature.

Motivated by the possibility of (de)activating patches, we introduce a new model of patchy colloids in which the patches can be either activated or deactivated by changing the temperature of the system. We find a very rich phase behaviour including lower critical points and reentrant percolation.

\section{Model and theory \label{model}}
\begin{figure}
\epsfig{file=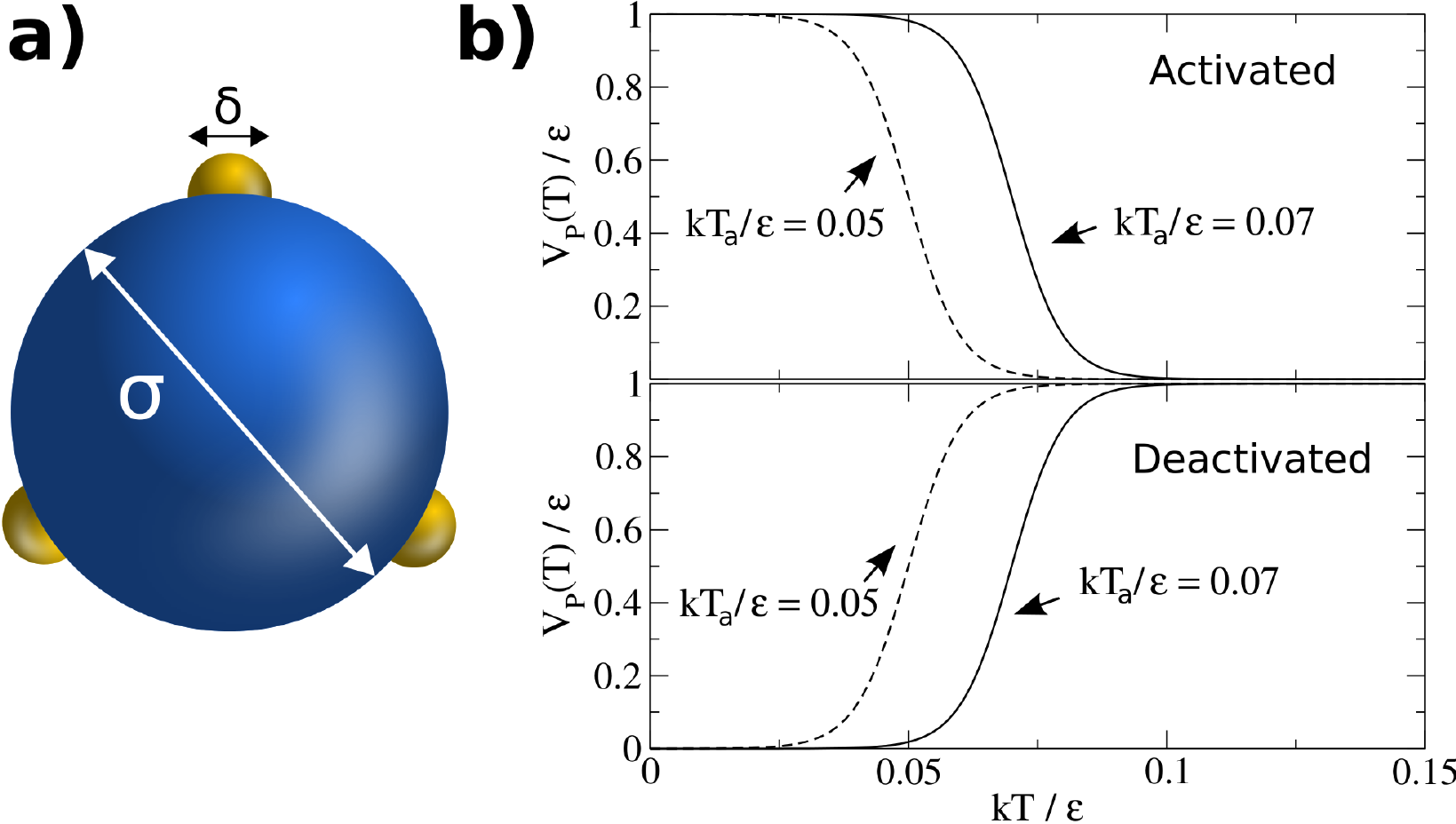,width=1.\linewidth,clip=}
\caption{(a) Schematic representation of the particles: hard spheres of diameter $\sigma=1$ with $N$ patches of diameter $\delta=0.119$ on the surface. (b) Interaction potential between two patches as a function of the temperature in the case of patches activated by temperature (top) and patches deactivated by temperature (bottom). In both cases we illustrate the potential for two activation temperatures $T_\text{a}$, as indicated. The width of the activation region is in all cases $\tau / \epsilon=0.01$.}
\label{fig1}
\end{figure}

The patchy colloids are modelled by hard spheres of diameter $\sigma=1$, representing a hard colloidal core, and small spheres of diameter $\delta$ located on the surface of the hard cores that play the role of bonding sites or patches, see Fig.~\ref{fig1}a. The patches can be either activated or deactivated by decreasing the temperature $T$. That is, two patches interact if they overlap (square well potential) and the temperature is below (temperature activated patches) or above (temperature deactivated patches) a given threshold. To model this behaviour we introduce a temperature dependent interaction potential between two patches:
\begin{equation}
V\text{p}(T)=
\begin{cases} 
\frac{\epsilon}{2}\left[1+\xi\tanh\left(\frac{T-T_\text{a}}{\tau}\right)\right], \text{patches overlap}  \\ 
0, \text{ patches do not overlap} 
\end{cases}
\label{potential}
\end{equation}
where $\epsilon=1$ sets the unit of energy, $T_\text{a}$ is the activation or deactivation temperature, and $\tau$ controls the width of the activation (or deactivation) region. The parameter $\xi$ is either $1$ or $-1$. In the former case the interaction between patches is deactivated by decreasing the temperature below $T_\text{a}$, whereas in the latter the interaction is activated by decreasing the temperature below $T_\text{a}$. Illustrative examples of the potential are shown in Fig.~\ref{fig1}b.

We use Wertheim's first order perturbation theory to study the thermodynamics of the system. Wertheim's theory has been successfully applied, as compared to computer simulations, to the same model in the case of "athermal" patches, that is patches for which the interaction potential is independent of the temperature. We, therefore, expect a similar level of agreement in the present case. A detailed description of Wertheim's thermodynamic perturbation theory for pure fluids can be found e.g. in Refs. \cite{wertheim1,*wertheim2,*wertheim3,*wertheim4,Chapman:1057}. The Helmholtz 
free energy per particle $f_\text{H}$ contains two contributions: the free energy of a reference system of hard spheres $f_\text{HS}$, and a perturbation accounting for the interactions between bonding sites $f_\text{b}$:
\begin{eqnarray}
f_\text{H}=F/N=f_\text{HS}+f_\text{b},
\end{eqnarray}
where $N$ is the total number of colloids. As usual, the free energy of a system of hard spheres can be expressed as the sum of the ideal-gas contribution and the excess part which describes the excluded volume effects: $f_\text{HS}=f_\text{id}+f_\text{ex}$. The ideal-gas free energy per particle is given exactly by
\begin{equation}
\beta f_{id}=\ln\eta-1
\end{equation}
where $\beta=(kT)^{-1}$ with $k$ the Boltzmann constant, and $\eta$ is the total packing fraction. $\eta=v_s\rho$, with $\rho$ the total number density and $v_s=\pi/6\sigma^3$ the volume of the hard sphere. In the above expression we have set the, irrelevant, thermal volume equal to $v_s$. The excess part accounts for the excluded volume between the hard cores of the colloids. We use the well-known Carnahan-Starling equation of state \cite{carnahan:635}:
\begin{equation}
\beta f_{ex}=\frac{4\eta-3\eta^2}{(1-\eta)^2}
\end{equation} 
Within Wertheim's first-order perturbation theory the bonding free energy per particle is given by \cite{Chapman:1057}
\begin{equation}
\beta f_b=N_\text{p}\left(\ln X-\frac{X}{2}+\frac12\right),\label{fb}
\end{equation}
where $N_\text{p}$ is the number of patches per colloid and $X$ is the probability that a bonding site is {\it not} bonded. The probability $X$ is related to the total density  through the mass-action equation:
\begin{equation}
X=\frac1{1+\eta N_\text{p}X\Delta}.\label{xnotbonded}
\end{equation}
The parameter $\Delta$ characterises the bond between two bonding sites, and can be written as
\begin{equation}
\Delta=\frac1{v_s}\int_{v_\text{b}}g_\text{HS}({\bf r})\left[\exp(\beta V_\text{p}(T))-1\right]d{\bf r},\label{delta}
\end{equation}
with $g_\text{HS}({\bf r})$ the pair correlation function of the reference fluid of hard spheres. The integral in the above expression is calculated over the bond volume $v_\text{b}$. Wertheim's theory assumes that each site can be involved only in a single bond and that two particles can form single bonds between them (no double bonds between two colloids are permitted). Both conditions are satisfied by geometrical constrains if one chooses a patch size $\delta=0.119$ which results in a bonding volume $v_b=0.000332285\sigma^3$, see \cite{Bianchi1} for details. Next, we approximate the pair correlation function by its contact value. Within this approximation
\begin{equation}
\Delta=\frac{v_b}{v_s}\left[\exp(\beta V_\text{p}(T))-1\right]A_0(\eta),\label{delta2}
\end{equation}
where
\begin{equation}
A_0(\eta)=\frac{1-\eta/2}{(1-\eta)^3}.
\end{equation}

In addition to the thermodynamics of the system, we also study the percolation behaviour. To this end we use the well-known Flory-Stockmayer \cite{flory1,*stock1,*flory2} theory of polymerization, according to which the system is percolated if the probability of finding a patch bonded, given by $1-X$, is higher than the threshold: 
\begin{equation}
p=\frac1{N_\text{p}-1}.
\end{equation}
The theory assumes a tree-like cluster structure with no closed loops. It is, therefore, consistent with Wertheim's first order perturbation theory, in which graphs with closed loops are not included in the Mayer expansion. The theory predicts a percolation threshold in good agreement with simulations for systems with temperature independent patches. We expect a similar level of agreement in the present model.

\section{Results \label{results}}
\subsection{Athermal patches}
\begin{figure}
\epsfig{file=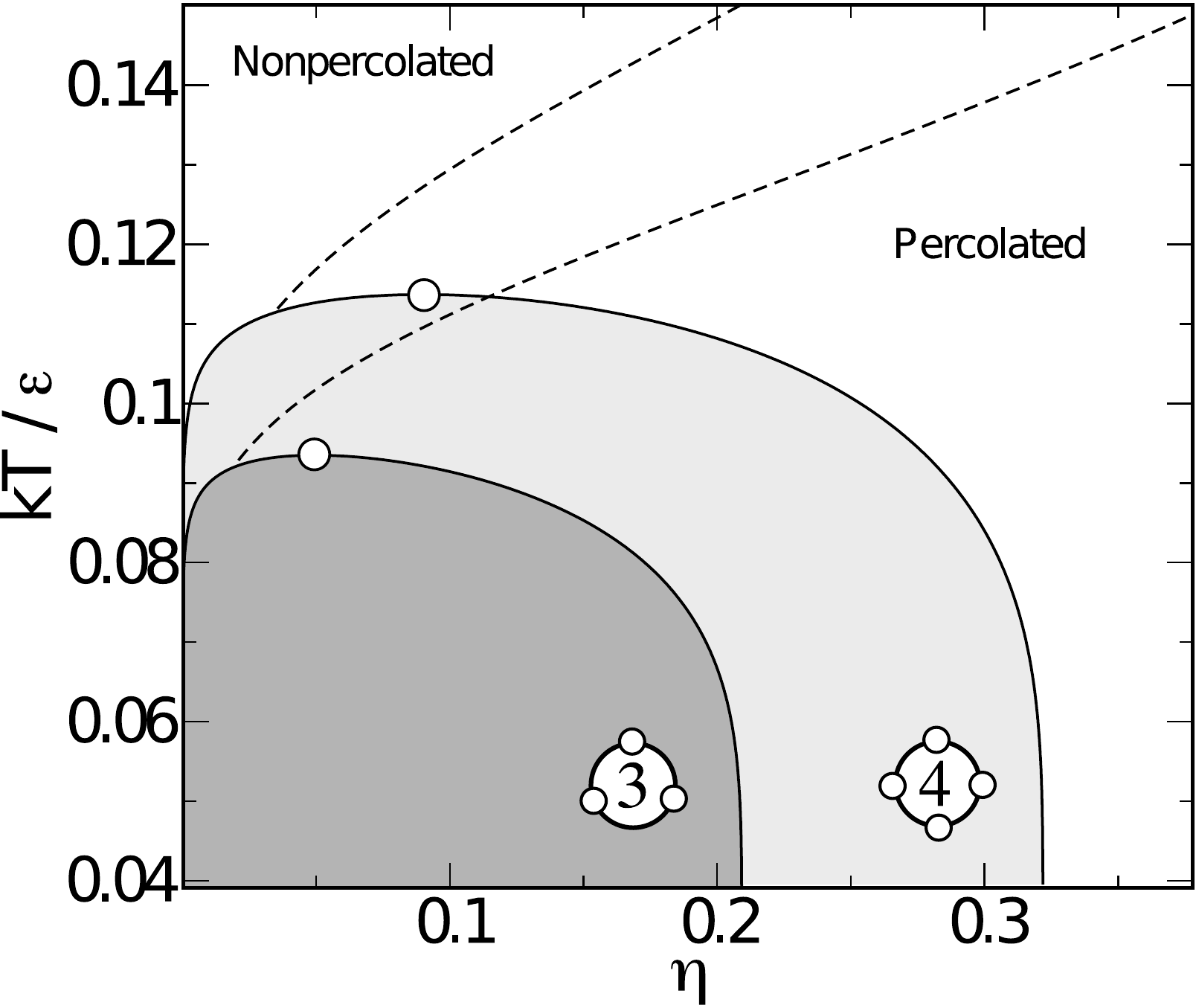,width=.9\linewidth,clip=}
\caption{Reduced temperature-packing fraction phase diagram of a single component fluid of hard-spheres with $3$ and $4$ athermal patches, as indicated. The solid curves are the binodal lines. The gray areas are the two-phase regions. The critical points are represented by empty circles. The dashed lines are the percolation lines.}
\label{fig2}
\end{figure}

We first show, as a reference, the behaviour of a system where all the patches are "athermal" in the sense that the interaction potential between two sites does not depend on the temperature. The energy of the system decreases by a fixed quantity $\epsilon$ whenever a bond is formed, independently of the temperature. This system has been extensively studied by theory and simulation \cite{Bianchi1}. The phase diagram is shown in Fig. \ref{fig2} in the plane temperature vs. packing fraction for particles with three and four patches. Below a critical temperature, the system exhibits coexistence between two fluids with different densities and fractions of bonded patches. The percolation line intercepts the binodal on the low-density phase. Hence, the high density phase is always percolated. It is often called a network fluid. The phase behaviour resembles that of "adhesive" hard spheres with isotropic short-range attraction~\cite{Re1}.

For particles with three or more patches the phase diagram is qualitatively the same. The main difference is the size of the two-phase region. It shrinks by decreasing the number of patches. Particles with only one patch tend to form dimers and fluid-fluid phase separation is absent. Particles with two patches tend to form chains. The absence of branching in this case prevents the fluid from phase separating.  

Next we study the case of thermal patches, with an interaction potential given by Eq. \eqref{potential}. 

\subsection{Thermally activated patches}
\begin{figure}
\epsfig{file=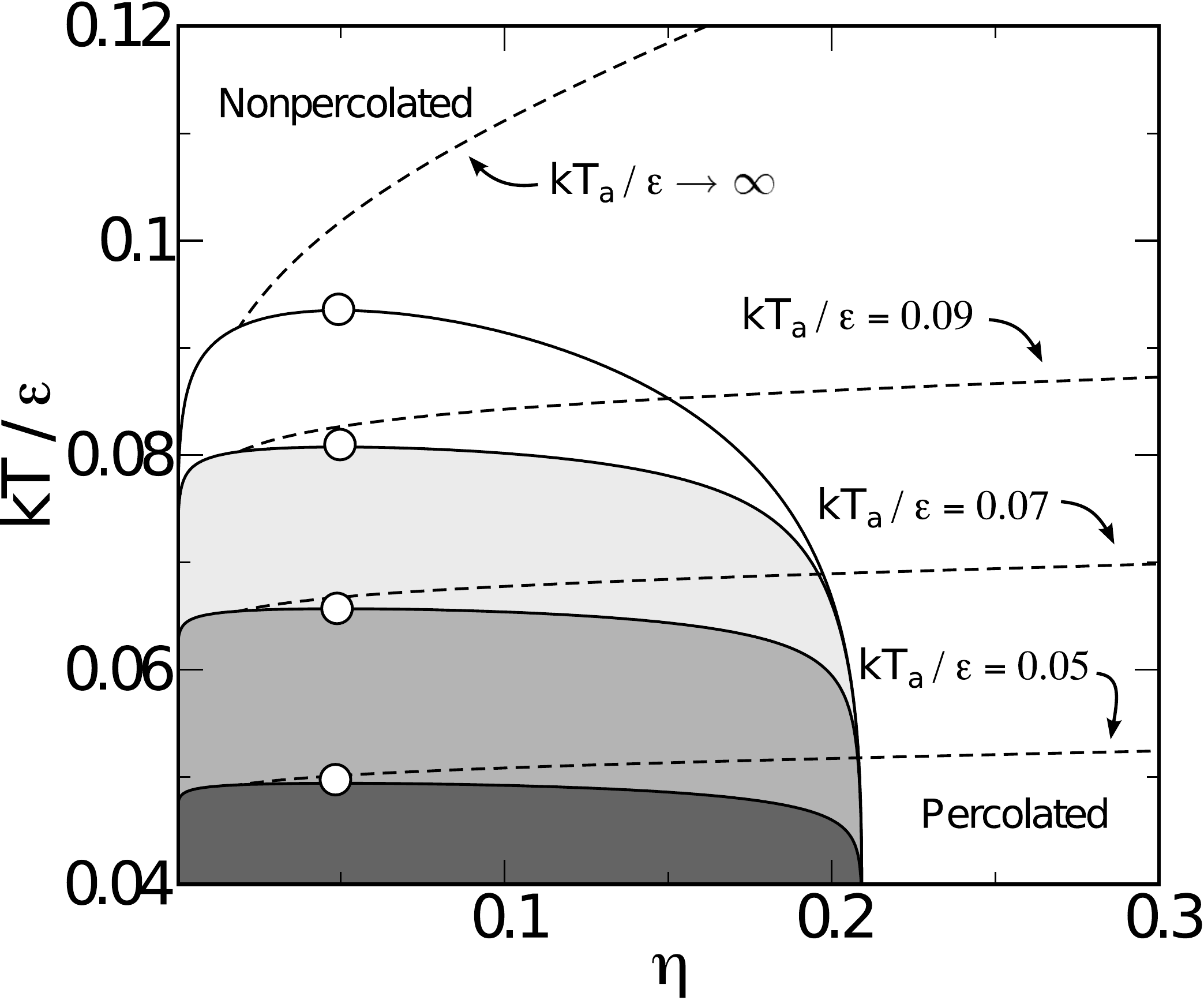,width=.9\linewidth,clip=}
\caption{Reduced temperature-packing fraction phase diagram of a single component fluid of hard-spheres with $3$ patches activated by temperature. The solid curves are the binodal lines. The gray areas are the two-phase regions. Dashed lines are the percolation lines. The critical points are represented by empty circles. Shown are four different cases corresponding to different activation temperatures: $kT_{a}/\epsilon\rightarrow\infty$ (athermal patches), $0.09$, $0.07,$ and $0.05$. In all cases the width of the activation region is $k\tau/\epsilon=0.01$. } 
\label{fig3}
\end{figure}

In Fig. \ref{fig3} we show the phase diagram (temperature-packing fraction plane) of a system with $3$ thermally activated patches with activation temperatures $kT_a/\epsilon=0.09$, $0.07$ and $0.05$. That is, the patches are active only at temperatures below the activation temperature. The width of the activation region is in all cases $k\tau/\epsilon=0.01$. As a reference, we also show the case of athermal patches ($kT_a / \epsilon\rightarrow\infty$). The topology of the phase diagram remains the same, independently of $T_\text{a}$. There is a phase coexistence between a low-density, mostly nonpercolated, fluid and a percolated, high-density fluid. The activation temperature determines the position of the critical point, which is located at temperatures slightly lower than $T_a$ (provided that $T_a$ is lower than the critical temperature of the system with athermal patches). The percolation line is also affected by the activation temperature, as no percolated states are present at temperatures above the activation temperature. 

\subsection{Thermally deactivated patches}

We depict in Fig. \ref{fig3} the temperature-packing fraction phase diagram of a system with three identical patches deactivated by temperature for two distinct deactivation temperatures: $kT_{\text{a}}/\epsilon,0.07,$ and $0.05$. Patches are active at temperatures below $T_{\text{a}}$. In both cases the deactivation temperature is below the critical temperature of a system with $3$ athermal patches, and the width of the activation region is $k\tau/\epsilon=0.01$. The phase diagram exhibits a closed two-phase region bounded by upper and lower critical points. The temperature of the lower critical point can be controlled by varying the deactivation temperature.

The system exhibits reentrant behaviour in both the high- and the low-density fluid phases. The percolation line collapses with the binodal close to both critical points, giving rise to a region of intermediate temperatures where the system is percolated. It is therefore possible to obtain the  sequence nonpercolated-percolated-nonpercolated by lowering the temperature at constant packing fraction.
\begin{figure}
\epsfig{file=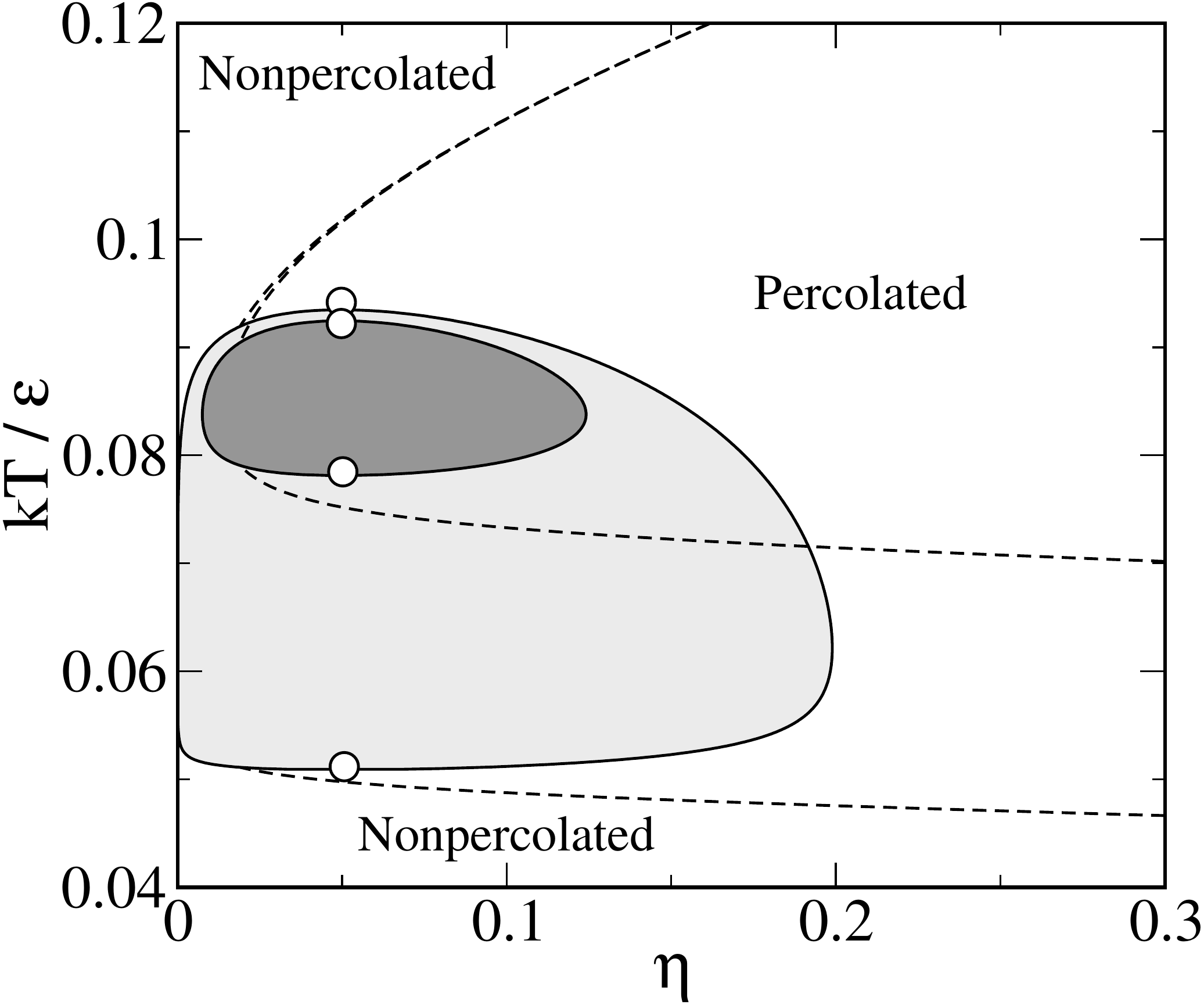,width=1.\linewidth,clip=}
\caption{Reduced temperature-packing fraction phase diagram of a single component fluid of hard-spheres with $3$ patches deactivated by temperature. The solid curves are the binodal lines. The gray areas are the two-phase regions. Dashed lines are the percolation lines. The critical points are represented by empty circles. The two different cases correspond to distinct deactivation temperatures: $kT_{\text{a}}/\epsilon=0.07,$ and $0.05$. In both cases $k\tau/\epsilon=0.01$. } 
\label{fig4}
\end{figure}

\subsection{Mixture of activated and deactivated patches}
\begin{figure}
\epsfig{file=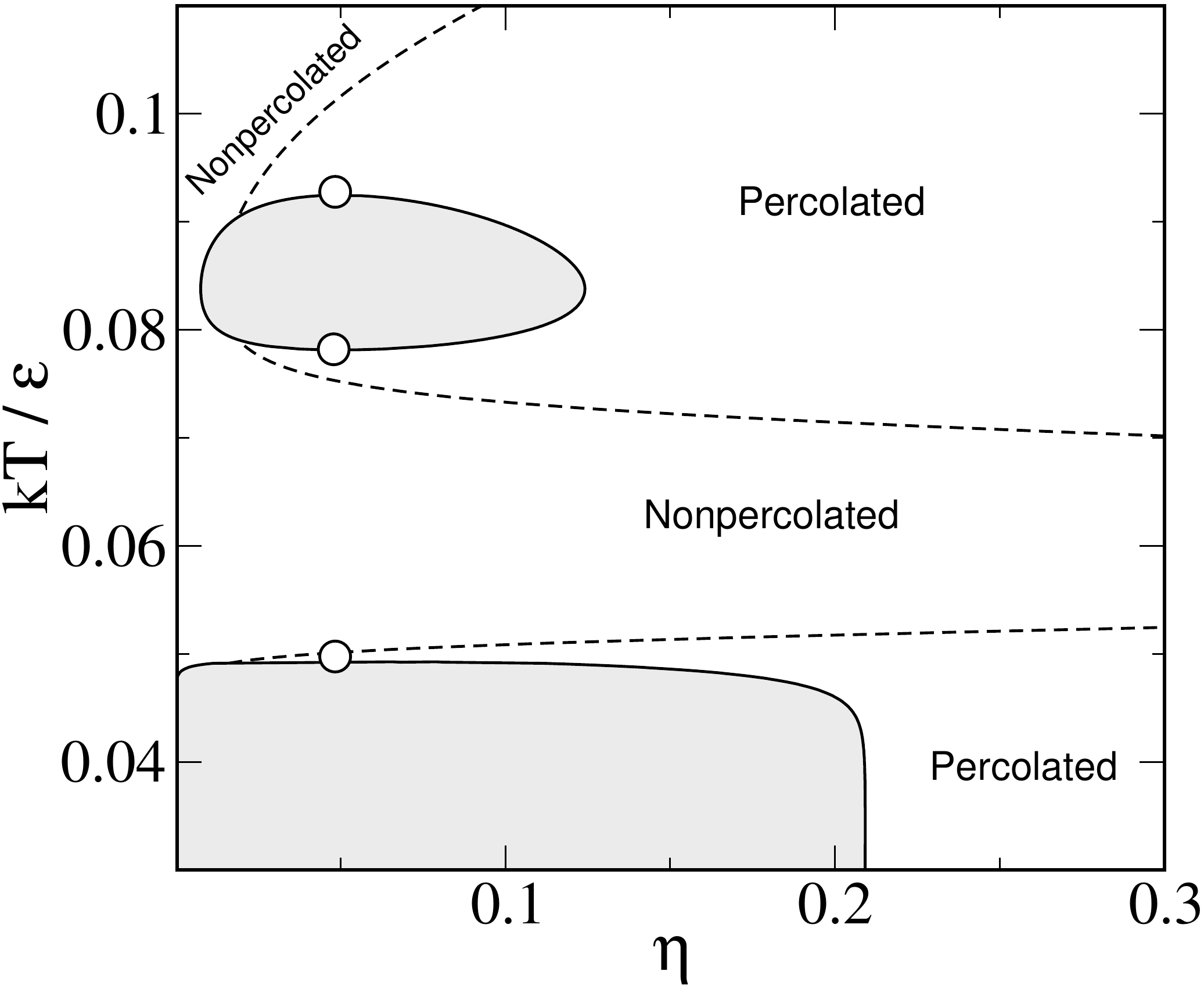,width=1.\linewidth,clip=}
\caption{Reduced temperature-packing fraction phase diagram of a single component fluid of hard-spheres with $3$ temperature activated $kT_\text{a}/\epsilon=0.05$ and $3$ temperature deactivated $kT_\text{a}/\epsilon=0.07$ patches. The solid curves are the binodal lines. The gray areas are the two-phase regions. Dashed lines are the percolation lines. The critical points are represented by empty circles. For both types of sites the activation width is $k\tau/\epsilon=0.01$. Only patches of the same type interact.} 
\label{fig5}
\end{figure}

Next we show an example in which each particle contains three temperature activated patches $kT_\text{a}/\epsilon=0.05$ and three temperature deactivated patches $kT_\text{a}/\epsilon=0.07$. To simplify the model only patches of the same type interact, that is, interactions between patches activated and deactivated by temperature are forbidden. The phase diagram is depicted in Fig. ~\ref{fig5}. Only three patches are activated at high temperatures, and only three are activated at low temperatures. There is a region, at intermediate temperatures, in which no patches are activated. As a consequence the phase diagram consists of two disconnected regions of phase separation. At high temperatures we find a closed loop of immiscibility bounded by upper and lower critical points, whereas at low temperatures there is a large two phase region bounded by an upper critical point. The high density fluid is percolated in the ranges of temperatures where the patches are activated. The intermediate region of temperatures, where no patches are activated, is nonpercolated. Hence, there is reentrant percolation, allowing the sequence percolated-nonpercolated-percolated-nonpercolated by increasing the temperature at fixed packing fraction.

\section{Discussion and conclusions}

We have focused on the bulk properties of a model with temperature dependent bonding sites. There are in the literature several works that extend Wertheim's theory to inhomogeneous situations. For instance, in \cite{Segura} Segura et al. used the weighted density functional theory of Tarazona \cite{PhysRevA.31.2672} to treat the excess part of the reference fluid of hard spheres, and extended the homogeneous bonding free-energy, cf. \eqref{fb}, by using:
\begin{equation}
\beta F_{\text{b}}[\rho({\bf r})] = N_{\text{p}} \int d{\bf r}\rho({\bf r})\left(\log X({\bf r})-\frac{X({\bf r})}2-\frac12\right)~\label{inhomogeneous}
\end{equation}
where $F_{\text{b}}$ is now a free-energy functional of the one body density $\rho({\bf r})$ which depends in general on the spatial coordinates ${\bf r}$. In \cite{YuWu} Yu and Wu treated the reference hard sphere fluid with the original Rosenfeld fundamental measure theory (FMT) \cite{PhysRevLett.63.980} and wrote the free energy of bonding in terms of the weighted densities of FMT. Both approaches were originally used to study a system of hard spheres with four associating sites in contact with a hard wall. In both cases there is a good agreement, even quantitative, with computer simulations. Similar approaches have been used to study e.g., the liquid-vapor interface \cite{Paulo} and the confinement in slit-pores \cite{Kalyu} of two models of patchy colloids that show reentrant phase behaviour in the bulk.

These extensions of Wertheim theory to inhomogeneous situations are a valuable tool and provide an accurate description of systems where the bonding sites are isotropically distributed, i.e. there is no preferential orientation of the particles. Sciortino et al. used both density functional theories to study hard spheres with three bonding sites in contact with a hard wall \cite{Nicoletta} and compared the results to Monte Carlo simulations. At high temperatures both approaches provide a good description of the system. At low temperatures, a regime where the energy of bonding dominates, the simulations show a desorption of particles close to the hard wall followed by an incipient layering structure. If the temperature is low enough the system decreases the free-energy by maximizing the number of bonds. As the wall is neutral, the particles next to the wall orient themselves moving their patches away from the neutral wall. As a result, the orientational order close to the wall is no longer isotropic. Both density functional approaches fail to describe the low temperature regime since the orientation of the particles is not taken into account by the theory. Therefore, simple extensions of Wertheim's theory to inhomogeneous situations, such as Eq. \eqref{inhomogeneous}, should be used with caution as they provide an accurate description only if the orientational order of the particles is irrelevant. In practice this may be case if the system under consideration meets at least one of the following criteria: (i) highly symmetric patch distributions (like four patches tetrahedrally distributed on the surface of a sphere), (ii) high temperatures, (iii) low densities where only fluid phases are involved.

In a recent work Marshall and Chapman \cite{Chapman1} have developed the first density functional that goes beyond the single bond condition of Wertheim's theory. The results are in excellent agreement with computer simulation. Telo da Gama et al. \cite{PhysRevE.88.060301} have developed a density functional that accounts for the orientations in the case of particles with two patches located on opposite poles of the spherical hard core. Also very recently, Marshall \cite{Marshall1} has been able to incorporate the orientational degrees of freedom in a one dimensional model of patchy particles with two patches. These recent achievements are promising towards the development of a density functional for patchy colloids that adequately describes the orientation of the patches.

We have studied the bulk phase behaviour of a new model of patchy colloids in which the bonding sites are either activated or deactivated below a given temperature. The model exhibits a rich phase behaviour. For example, for a system with three patches deactivated by temperature the two phase region ends at a lower critical point, a very unusual feature in single component systems. Recently, a very similar phase diagram has been predicted for a model of patchy particles \cite{PhysRevLett.111.168302,C5SM00559K} with two types of patches. The interaction energies and the number of patches of each type can be tuned in such a way that at low temperatures the system favours the formation of isolated rings, which eventually leads to a lower critical point. In our case the lower critical point is associated to the deactivation of the patches that induces the formation of single monomers. Therefore both mechanisms, although different, give rise to a lower critical point by breaking the connectivity of the system. 

For particles with both types of patches, activated and deactivated by temperature, we have found an even more unusual behaviour. The phase diagram shows a closed loop bounded by two critical points at high temperature, and a low temperature demixing region bounded by an upper critical point. A similar phase diagram has been predicted in a binary mixture of patchy colloids \cite{heras:104904}, but not in single component systems as reported here.

We have also investigated (not shown) more complex mixtures, involving e.g., thermal and athermal patches. By adjusting the types and number of patches one can basically sculpt the shape of the coexistence curve between percolated and nonpercolated phases. 

We have focused on the low density regime, where only fluid phases are stable. At higher packing fractions solid phases will be stable. If the patches are deactivated or the temperature is high enough the stable crystal structure will be that of pure hard spheres. On the other hand, if the patches are activated they will play a major role on determining the stable crystal structure. For instance, by varying the size of the patches in particles with four bonding sites, it is possible to stabilize diamond or bcc structures \cite{Sciortino2}. Our model can be used to study the interplay between the entropically driven crystal phase of hard spheres and the crystal structures formed by patchy colloids.

\section{Acknowledgement}
This work has been partially funded by the Portuguese Foundation for Science and Technology (FCT) through project EXCL/fis-nan/0083/2012.

\end{document}